\documentclass[%
 amsmath,amssymb,
 aps, physrev,
 twocolumn,
]{revtex4-2}

\usepackage{graphicx}
\usepackage{dcolumn}
\usepackage{bm}
\usepackage[dvipsnames]{xcolor}
\usepackage{makecell}

\usepackage{tabularray}
\UseTblrLibrary{diagbox} 

\begin{document}

\preprint{APS/123-QED}

\title{\textbf{Chiral cellulose fibril organization in a plant cell wall as liquid crystal confined in a cylindrical boundary} 
}%

\author{Jieh-Wen Tsung}
 \email{Contact author: jiehwen.tsung@nycu.edu.tw}
\author{Bo-Hsien Wu}%
\author{Li-Yan Hung}%
\author{Yu-Chieh Huang}%
\author{Yueh-Ching Huang}%
\author{Shao-Chun Hsieh}%
\affiliation{%
 Department of Electrophysics, National Yang Ming Chiao Tung University, Hsinchu, 300110, Taiwan\\
}%



\date{\today}

\begin{abstract}
The order of cellulose microfibrils in xylem cell walls is identified using polarized optical microscopy.
The line, helix, ring, crossed helix, and twisted helix organizations in a cylindrical shell are the equilibrium states of balanced elastic deformation and surface anchoring. A computational model of the five organizations is established to simulate their birefringent-colored profiles. These five textures present distinct, unique optical textures, which are clearly recognized in the cross sections of \textit{Eucalyptus grandis}. The libriform fiber cell for support has a twisted helical cell wall. The vessel cell for high-speed water uptake consists of a helical layer covered by a layer of vortex array. The ray cell for radial transportation is crossed helical. The microfibril angles versus the radius of the cell wall were measured utilizing the distribution of birefringence colors. In fiber cell walls, the MFA is significantly correlated with $\frac{R}{k_{33}/\gamma}$, where $R$, $k_{33}$, and $\gamma$ represent the curvature, bending, and surface anchoring, respectively. In vessel cell walls, the vortex array includes focal conic domains of chiral order and topological defects of nematic order, suggesting that the phase transition of cellulose fibrils leads to pattern formation. Liquid crystal phases and patterns in the cell walls reveal how the cell wall thickens and how cells differentiate. 
Out of the frustration of long, stiff, twisting fibrils packed in slender tubes, trees generate the helical channel networks, transforming the brittle lamina into an elastic, tear-resistant, self-healing tissue.
Beyond expectation, xylem cell walls preserve the patterns of phase transitions and spontaneous symmetry breaking. In analogy with the early universe, superconductors, magnetic moments, and liquid crystals, wood represents a biological system that utilizes the exact same physical laws for the sake of growth and form, which is tangible and accessible.

\end{abstract}

\maketitle


\section{\label{sec:level1}Introduction}

Trees are incredibly self-sustaining biological systems powered by solar energy to create their own structural materials. The product is wood, a robust, stable biopolymer composite that has been utilized for buildings, ships, tools, and papermaking for thousands of years. The cellulose derivatives extracted from the pulp are processed into pharmaceuticals and optoelectronics.  If humans steward the forests with long-term care, consuming the part we need and avoiding overuse, a living forest can be the most abundant, inexhaustible supply of material and energy.

To maximize solar energy collection, trees need tall, robust trunks with extending, flexible branches that can sway with the wind to hold a vast canopy of leaves. The trunk, branches, and leaf veins contain vascular bundles. Driven by leaf transpiration and capillary force within the vascular bundles, trees transport enormous volumes of water upward without an external pump. Consequently, these conduits must have high structural integrity to resist buckling or collapsing under the massive negative pressure generated by water uptake. Wood successfully satisfies these competing mechanical requirements simultaneously due to the hierarchical, chiral structure of its cellulose fibrils.

Decades ago, physicists studying biological pattern formation and liquid crystals discovered that cellulose assembles into chiral structures spontaneously~\cite{CLC_vivo_Bouligand, CLC_fibril_Bouligand, Plant_cell_LC_Neville_1993, CLC_living_Mitov, CLC_living_Godinho}.
The distribution of hydrogen bonds on a cellulose molecule is asymmetric, resulting in helical fibril bundles~\cite{CLC_helical_Godinho}. The fibril bundles align in order, embedded in a matrix of lignin and  hemicellulose. The membranes of aligned fibrils rotate progressively through the thickness, creating a twisted plywood architecture (Bouligand structure).
Similar twisted arrangements occur in collagen and chitin, meaning that these chiral structures exist in a wide variety of materials across plants, insects, and animals. This ubiquity strongly suggests that the formation of twisted plywood structure is a universal physical phenomenon. While genes and chemical reactions undoubtedly determine the structure of the molecules, the following physics of liquid crystal phase transitions and self-assembly likely drives and regulates macroscopic growth and form~\cite{thompson1942ongrowth}.

\begin{table*}
\caption{\label{table:intro_5helix}Cellulose microfibril structure of xylem cell walls and the key parameters of ordering. Elastic constants are $k_{11}$, $k_{22}$, and $k_{33}$ for splay, twist, and bend, respectively. R, radius of the cylinder. $l_0$, extrapolation length or internal length. Twisted helix is also called Bouligand Structure.}
\begin{ruledtabular}
\renewcommand{\arraystretch}{2}
\begin{tabular}{cccccc}
\diagbox[innerwidth = 3cm, height = 20ex]{}{MFA texture}
&\begin{minipage}{0.13\textwidth}
      \includegraphics[width=\linewidth]{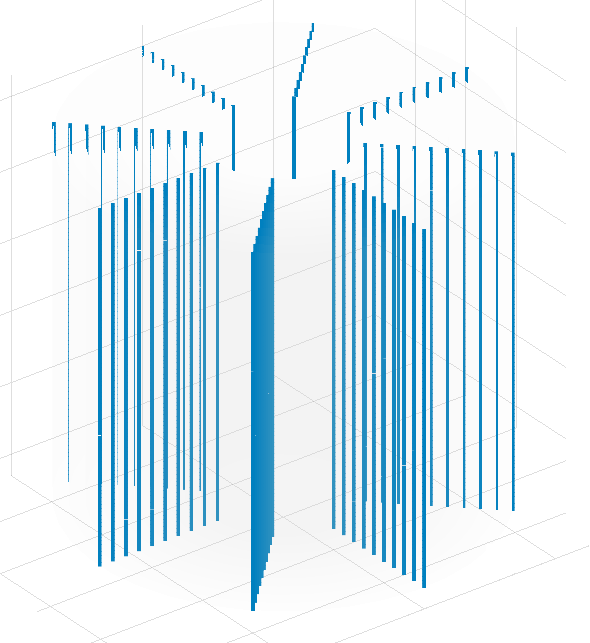}
    \end{minipage}
&\begin{minipage}{0.13\textwidth}
      \includegraphics[width=\linewidth]{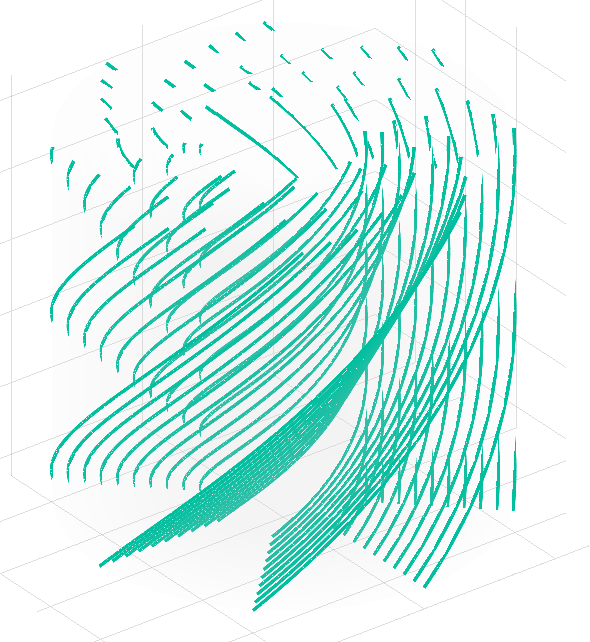}
    \end{minipage}
&\begin{minipage}{0.13\textwidth}
      \includegraphics[width=\linewidth]{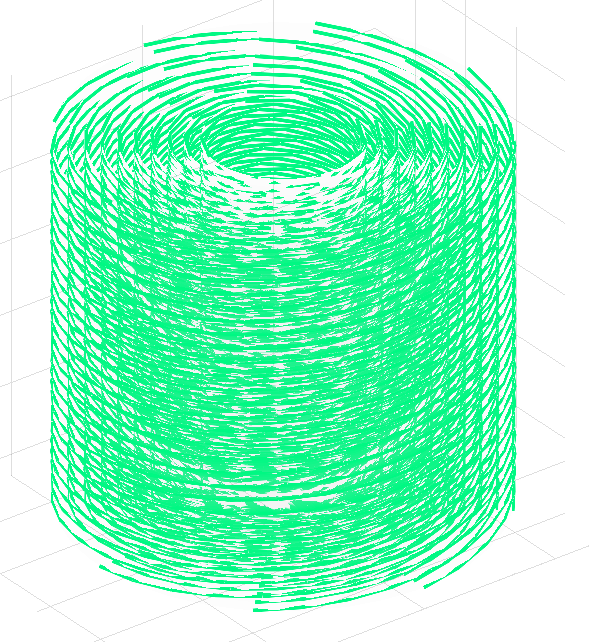}
    \end{minipage}
&\begin{minipage}{0.13\textwidth}
      \includegraphics[width=\linewidth]{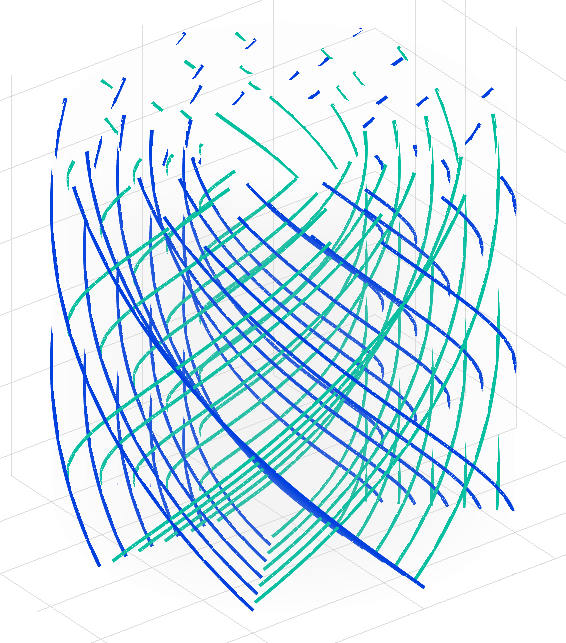}
    \end{minipage}
&\begin{minipage}{0.13\textwidth}
      \includegraphics[width=\linewidth]{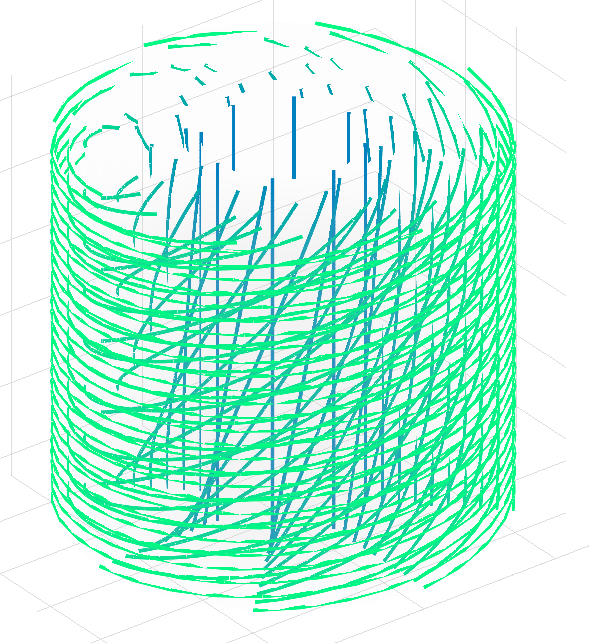}
    \end{minipage} \\
\textbf{Key parameters}
& \textbf{Line}
& \textbf{Helix}
& \textbf{Ring}
& \textbf{Crossed helix}
& \textbf{Twisted helix}\\
\hline
Order
& Nematic
& Nematic
& Nematic
& Nematic
& Nematic or chiral\\
\hline
Anisotropic elasticity 
& $\frac{k_{33}}{k_{11}}>1$
& 
& $\frac{k_{33}}{k_{11}}<0.5$
& 
& $k_{33} \geq k_{11} \gg  k_{22}$\\
\hline
Boundary (NLC~\cite{NLC_stable_Burylov})
& $\frac{R}{l_{0, k_{11}}} < 2 $
& 
& $\frac{R}{l_{0, k_{11}}} > 5$
& 
& $\frac{R}{l_{0, k_{11}}} > 7$\\
\hline
Boundary (fibril~\cite{NLC_stable_Rey})
& $\frac{R}{l_{0}} < 0.2$
& $0.2< \frac{R}{l_{0}} < 0.4$
& $\frac{R}{l_{0}} > 0.4$
& 
& \\
\hline
Mechanical benefits
& \makecell[c]{Stiffness}
& \makecell[c]{Preventing buckling}
& \makecell[c]{Hooping}
& \makecell[c]{Flexibility}
& \makecell[c]{Impact-resistance}\\
\hline
Examples
& \makecell[c]{Tension wood}
& \makecell[c]{S2 \\ of xylem Cells}
& \makecell[c]{S1 S3 \\ of xylem Cells}
& \makecell[c]{Grass \\ Conifer}
& \makecell[c]{Stone cell \\ Seed coat}\\
\hline
Reference
& \cite{MFA_review_Barnett}\cite{Helix_CrossHelix_Eder}
& \cite{MFA_review_Barnett}\cite{Helix_CrossHelix_Eder}
& \cite{MFA_review_Barnett}\cite{Helix_CrossHelix_Eder}
& \cite{CLC_nascent_Reis}\cite{CrossedHelix_Bamboo}
& \cite{Plant_cell_LC_Neville_1993}\cite{twistedHelix_seed_shell}\\
\end{tabular}
\end{ruledtabular}
\end{table*}

The orientation of a cellulose microfibril in the cell wall relative to the longitudinal fiber axis is defined as the microfibril angle (MFA). Five typical MFA arrangements were found in plants: line, helix, ring, crossed helix, and twisted helix (helicoid)~\cite{Rey_five_helix_2015, Rey_five_helix_2022}.
The schematic diagrams of the five structures are displayed in Table~\ref{table:intro_5helix}.
For the line, helix, and ring, the MFA is longitudinal, helical, and azimuthal with respect to the fiber axis, respectively.
Crossed helix means that the fiber rotates to the right and then to the left, layer-by-layer, alternately.
In the twisted helical tube, the microfibrils rotate from the axial (inside) to the tangential direction (outside) of the tube.
The mechanical benefits and examples of each texture are summarized in Table~\ref{table:intro_5helix}.


In most of the angiosperms (flowering plants), the xylem is composed of a radial system of ray cells and axial systems of vessel and libriform fiber cells. The ray cell transports water radially and store nutrients. The vessels are for high speed water uptake. The fiber cells supports the other cells, keeping the tree straight and upright.
A cell wall comprises primary, S1, S2,  and S3 layers~\cite{MFA_review_Barnett}. The primary, S1, and S3 are very thin. The MFA structures are significant in the thickest S2 layer.
Identifying the MFA structure of ray, vessel, and fiber cells is crucial for understanding how the tree grows tall, and how the enormous amount of water can be lifted upwards without expending energy.
For biological studies, the process of xylem development, especially, how the newborn cells differentiate and divide the labor, still remains mysterious.
The cause, timing, and type of helix formation can be very crucial for answering how the plant cells differentiate.

Scanning Electron Microscope (SEM), Transmission Electron Microscope (TEM), Atomic Force Microscope (AFM), and Confocal Laser Scanning Microscope (CLSM) have been utilized to study the cell wall structure~\cite{MFA_review_Barnett, Microscopy_for_Wood}. The resolution of these techniques is sufficient to show the cellulose fibrils, but the actual image is often limited to recognizing the S1, S2, and S3 layers~\cite{SEM_fiber_vessel_ray}. It is very difficult to find the fibrils twisting, since they are embedded in the lignin and hemicellulose matrix. Additionally, the dehydration, sputter coating, or cryo-fixation could have destroyed the helical microstructure.
To overcome the complications and damages, we apply polarized optical microscopy (POM), since the cellulose fibrils are birefringent but the matrix is optically isotropic.

POM is capable of showing the orientation of a cellulose fiber. Light through the birefringent material decomposes into ordinary and extraordinary modes that propagate at different velocities. The optical path/phase difference between the two modes is visualized as birefringent colors with polarizers, analyzers, and retardation plates under a POM~\cite{POM_Bloss, POM_Delly}. The color is related to the birefringence and the thickness of the sample, and the brightness is determined by the orientation of the fiber axis versus the polarizers. Therefore, without breaking, freezing, drying, or coating the plant tissue, the orientation of the cellulose fibrils can be clearly displayed as colors in the microscopic image. 
POM has been applied to measure the MFAs of the stem of wheat~\cite{MFA_POM_SEM}, which is a grass with simple helical cellulose organization. MFA measured using POM, SEM, and x-ray diffraction agree with each other well, meaning that POM is reliable.
This research provides a major improvement over previous methods.
We theoretically calculate the birefringence color of line, helix, ring, crossed helix, and twisted helix organizations under POM. The five structures should have their own distinct birefringence colors and optical textures. By recognizing the optical textures, the cellulose structure can be easily identified, and the birefringence color spectrum provides accurate MFA.

This article begins by identifying the cellulose microfibril structure of xylem cell walls.
Next, the liquid crystal ordering of the microfibrils and the conditions of their stable configurations are carefully verified.
Finally, we explore whether liquid crystal phase transitions are actively involved in cell wall development.
Beyond our expectation, evidence of ordering induced by geometrical frustration and liquid crystal phase transitions emerged under the POM. Long, stiff cellulose fibrils swirl to fit into a narrow cylindrical cell wall. Patterns of topological defects associated with nematic and chiral phase transitions appear on the vessel cell wall. Cellulose in wood could be the same as cosmological fields in the early universe, superconducting electron systems, magnetic moments, and liquid crystals, presenting the patterns of spontaneous symmetry breaking. While the patterns in superconductors, magnets, and liquid crystals can be transient and fleeting, wood preserves the history of phase transitions like a solidified universe, which is tangible and accessible.

In Literature Review, the theoretical models of liquid crystal ordering in a curved confinement are analyzed to find the key parameters.
The Model section demonstrates the computation program of the five typical cell wall structures, showing how the birefringence color of light through the cell walls is simulated. Both the transverse and longitudinal cross sections were simulated.
The Experimental Results start with the transverse and longitudinal slices of wood under POM, showing the birefringent-colored optical textures of the cells. The actual and simulated optical textures are compared to identify the microfibril structure.
The MFA data measured with the birefringence color spectrum are analyzed statistically.
In the Discussion section, two topics are explored: (a) How the competing curvature and surface anchoring ($\frac{R}{l_0}$) regulate the microfibril order, and (b) The evidence of liquid crystal phase transition during cell wall formation.

\section{\label{sec:literature}Literature review}

The cellulose fibrils align in order to have a dense packing, which leads to minimum free energy. The nascent cellulose fibrils align into chiral nematic order~\cite{CLC_vivo_Roland, CLC_nascent_Reis, CLC_vivo_Rey} before they are fixed in the matrix.
The cellulose fiber created by bacteria is a nematic fluid~\cite{CLC_vivo_Smalyukh}.
In a curved boundary, ordered, twisted fibrils result in minimum deformation and least free energy.
The line, helix, ring, crossed helix, and twisted helix can be the static equilibrium of elastic deformation and surface anchoring.
The key parameters are summarized in Table~\ref{table:intro_5helix}.

\subsection{Curvature-induced microfibril organization}
A mechanical model of the plant cell wall was developed assuming that a diluted dispersion of cellulose fibrils was evenly distributed on a curved elastic membrane~\cite{NLC_stable_Rey}. The model integrated the elastic energy of the membrane, the Landau-de Gennes fibril orientation energy, and the energy of curvophilic and curvophobic interactions between the fibril and the membrane. The curvophobic over the curvophilic energy would be an internal length scale ($l_{0}$). The radius of the cylinder R over $l_{0}$ determined the fibril structure.
$\frac{R}{l_0} < 0.2$ led to the line structure.
$0.2<\frac{R}{l_0} < 0.4$ was the range for a helix.
When $\frac{R}{l_0} > 0.4$, the fibrils followed the circular planar surface anchoring.
A list of possible curvophilic and curvophobic mechanisms were proposed with the model. However, how the microfibrils were resisting or attaching to the curved surface was not clearly stated.
This will be the focus of our experiment: looking for the curvophilic and curvophobic behavior of microfibrils in the cell wall.

\subsection{Elongation-induced helical structure}

Another mechanical model stated that the twisted helix is due neither to the guidance (cortical microtubules) nor to liquid crystal. It is simply because the cell was elongating during cellulose deposition~\cite{plant_cell_strech_Mulder}.
Cellulose synthesis complexes (rosettes) in the epidermal cells (plasma) generated the cellulose fibrils.
The fibrils deposited along the circumferential direction.
These azimuthal rings constrained the radial expansion, so the cell grew longer in the axial direction, pulling the microfibrils to reorient from the tangential to the axial direction.
Then the new layer of cellulose deposited on the outside of the old, stretched layer, pointing to the circumference direction again.
As the cell deposited layers of cellulose progressively over time, the MFA appears rotate consistently from the previous layer, building a twisted helical texture.

\subsection{Stable states of nematic liquid crystal in a cylinder}

Biophysicist, Y. Bouligand, and biologist, A. C. Neville, established that biological molecules behave as liquid crystals, based on the discovery of twisted plywood architectures in living tissues. The twisted plywood looks like a piece of solidified chiral liquid crystal~\cite{CLC_vivo_Bouligand, Plant_cell_LC_Neville_1993}.
They noted that concentrated solutions of purified biopolymers, such as chitin, collagen, cellulose, and DNA, spontaneously form liquid crystal phases. The alignment relies on the rigid-rod nature of these molecules.
It has been long hypothesized that the periodic array of tubes and vortices in bones, beetle shells, and plant cell walls can be the patterns of self-assembled liquid crystal.

The equilibrium configuration of nematic liquid crystal (NLC) confined in a cylinder is determined by the anisotropic elastic deformation and how the NLC is anchored on the boundary surface (vertical or in-plane)~\cite{NLC_stable_Cladis_Kleman, NLC_stable_Burylov}.
Splay, twist, and bend in NLC can have various elastic constants, which are $k_{11}$, $k_{22}$, and $k_{33}$, respectively. Cellulose fibers are long and stiff. The hydrogen bond pulls them closer, while the asymmetric distribution of the bonds introduce chirality. Therefore, the cellulose in water has $k_{33} \geq k_{11} \gg  k_{22}$, meaning that cellulose resists bending and splaying but tend to twist to fit into a slender cylinder~\cite{NLC_stable_twist_Yodh}. The axial, planar circular, and escaped circular configuration in a cylinder with the circularly in-plane surface anchoring are similar to the line, ring, and twisted helix, respectively. 

In previous works~\cite{NLC_stable_Cladis_Kleman, NLC_stable_Burylov},
the key parameters were $\frac{k_{33}}{k_{11}}$ and $\frac{R}{l_0}$, where $R$ was the radius of the cylinder and $l_0$ was the extrapolation length. 
Extrapolation length is the distance away from the boundary, where the anchoring becomes weak and negligible, defined as the elastic constant over anchoring strength.
In the work by Burylov~\cite{NLC_stable_Burylov}, $l_{0, k_{11}}$ equaled $\frac{k_{11}}{W_{\theta}}$, measuring how easy the NLC could lift itself from the surface, where $k_{11}$ was for splay deformation and $W_{\theta}$ was the anchoring strength of the out-of-plane perturbations. 
The line structures were stiff NLC ($\frac{k_{33}}{k_{11}}>1$) packed in a narrow tube with a strong anchoring ($\frac{R}{l_{0, k_{11}}} \lesssim 2 $).
The rings were soft NLC ($\frac{k_{33}}{k_{11}}<0.5$) aligned in a wide tube with a weak anchoring ($\frac{R}{l_{00, k_{11}}} \gtrsim 5$).
The stiff NLC in a wide cylinder preferred the circular planar polar structure of two topological defects attached on the wall
when $ 0.5 < \frac{k_{33}}{k_{11}} < 3.2$ and $ 2 < \frac{R}{l_{0, k_{11}}} < 7$.
In a very wide cylinder with weak anchoring ($ 7 < \frac{R}{l_{0, k_{11}}}$), the NLC ($ 0.5 < \frac{k_{33}}{k_{11}}$) could lift itself from the surface and twisted to escape at the middle of the cylinder, leading to the twisting escaped circular structure.
However, later work by Bates~\cite{CLC_stable_Bates} reported that the twisted achiral NLC was not stable in a cylindrical confinement. Instead, the NLC relaxed into untwisted states to achieve the minimum free energy. Introducing chirality was necessary to maintain the twisted structures in a cylinder.
The chiral NLC aligned along the axis at the center of the cylinder, and twists away versus the increasing thickness.
The crossed helix is not directly related to any LC stable states. It can be short-pitched CLC, isotropic state, or a result of fast growing~\cite{CLC_nascent_Reis}.

\subsection{Elasticity and interfacial tension of cellulose nanocrystal}
Cellulose nanocrystals are rod-like colloidal particles in water, showing liquid crystal phases.
The isotropic, nematic, and chiral phases can coexist but separate. The droplets of the chiral parts (tactoids) coalesce, precipitate, and form an ordered lamina. The transition volume of the droplet is the result of balanced elastic deformation and surface tension. Thus the extrapolation length ($l_0$), which is the elastic constant over surface anchoring energy, was successfully measured in vitro~\cite{CNC_Extra_length_Bagnani}. Here the surface tension coefficient was $\gamma$, weighted with $\omega_\theta$ or $\omega_\phi$ if the tension was anisotropic. The sulfated cellulose nanocrystal in water showed isotropic and uniaxial chiral phases with $\frac{k_{11}}{\gamma}= 11 \pm 9 $~\textmu m, $\frac{k_{22}}{\gamma}= 0.09 \pm 0.05 $~\textmu m, and $\frac{k_{33}}{\gamma}= 9.6 \pm 1.5 $~\textmu m.
The carboxylated cellulose nanocrystal in water showed isotropic, nematic, and uniaxial chiral phases with 
$\frac{k_{11}}{\gamma}= 5 \pm 3 $~\textmu m, $\frac{k_{22}}{\gamma}= 0.8 \pm 0.1 $~\textmu m, and $\frac{k_{33}}{\gamma}= 8 \pm 0.1 $~\textmu m.
$l_{0, k_{11}}$ and $l_{0, k_{33}}$ are about the size of the a plant cell, meaning that the interfacial tension is strong enough to align the cellulose deposited on the cell wall.
The $k_{22}$ was about $\frac{1}{10}$ of the $k_{11}$ and $k_{33}$, meaning that the fibrils prefer twisting in a packed circular boundary.

In analogy to the liquid crystal confined in a capillary tube or in a droplet, a rigid rod-like fiber with large $k_{33}$ is curvophobic, while a circular planar surface alignment due to the interfacial tension (hoop stress) can make the fibrils curvophillic.
The $l_{0}$ in the mechanical model is likely to be the extrapolation length. In this research, we aim to verify the $l_{0, k_{33}} = \frac{k_{33}}{\gamma \omega_{\phi}}$ relation with real wood tissue.

\section{\label{sec:model}Model}

\subsection{Birefringence color showing the orientation of a fiber}
Cellulose fibrils are birefringent.
Light through the birefringent material decomposes into ordinary and extraordinary modes, of which the refractive indices are $n_o$ and $n_e$, respectively. The two modes travel at different speeds, resulting in an optical path/phase difference (OPD) that changes the polarization state of the light. Linearly polarized light passing through the birefringent material is converted into elliptically polarized light. OPD is a function of the birefringence ($\Delta n = n_{eff} - n_o$), the thickness of the sample ($d$), and the wavelength of the light ($\lambda$):
\begin{equation} \label{eq:OPD}
OPD = \sum_{i} (n_{eff} - n_o) \cdot d_i
\end{equation}
\begin{equation} \label{eq:n_eff}
n_{eff} = \frac{n_o n_e}{\sqrt{(n_e\cos\alpha)^2 + (n_o\sin\alpha)^2}}
\end{equation}
where the angle $\alpha$ is between the direction of light propagation ($\hat k$) and the optical axis of the fibril ($\hat c$). The orientation of the fibril determines the OPD.

Under a POM, the sample is placed between the crossed linear polarizer and analyzer. The polarization state of light ($\vec{E}_{i} = E_{ix} \hat x + E_{iy} \hat y$) changes as
\begin{align*}
    &\begin{bmatrix}
        E_{ox} \\
        E_{oy} \\
    \end{bmatrix} \\
    &=
    \left[ {\begin{array}{cc}
    \cos{\beta}  &-\sin{\beta} \\
    \sin{\beta}  &\cos{\beta} \\
    \end{array} } \right]
    \left[ {\begin{array}{cc}
    e^{-i\frac{\Gamma}{2}} & 0 \\
    0                      & e^{+i\frac{\Gamma}{2}} \\
    \end{array} } \right]
    \left[ {\begin{array}{cc}
    \cos{\beta}  &\sin{\beta} \\
    -\sin{\beta}  &\cos{\beta} \\
    \end{array} } \right]
    \begin{bmatrix}
        E_{ix} \\
        E_{iy} \\
    \end{bmatrix} \\
    & = \begin{bmatrix}
        \sin{\beta}\cos{\beta} \left(e^{-i\frac{\Gamma}{2}} - e^{+i\frac{\Gamma}{2}} \right) \\
        \sin^2{\beta}e^{-i\frac{\Gamma}{2}} + \cos^2{\beta}e^{+i\frac{\Gamma}{2}} \\
    \end{bmatrix}
     \begin{bmatrix}
        E_{ix} \\
        E_{iy} \\
    \end{bmatrix}
\end{align*}
where $\Gamma$ is the optical phase retardation, $2 \pi \frac{OPD}{\lambda}$.
$\beta$ is the angle between
the projection of $\hat c$ on the polarization plane of the light (the plane perpendicular to $\hat k$, where the electric and magnetic fields oscillate) and the transmission axis of the polarizer.
The intensity of the output light ($\vec{E}_{o} = E_{ox} \hat x + E_{oy} \hat y$) is
\begin{equation} \label{eq:I_bi_color}
\frac{I}{I_0} = \frac{\mid {E_{ox}} \mid^2}{ \mid {E_{ix}} \mid^2 + \mid {E_{iy}} \mid^2 } =\sin ^2{ \left( 2\beta \right) } \sin^2{  \left(\pi\frac{OPD}{\lambda} \right) }
\end{equation}
The POM image displays birefringence colors.
The color ($\lambda$) is related to the OPD and thus the polar orientation of the fibril ($\alpha$), see the birefringence color chart in Fig.~\ref{fig:Xylem_MFA_5modes_opt_top}.
The brightness shows the azimuthal orientation of the fibrils ($\beta$).

\subsection{Simulation of the five cell walls under polarized optical microscope (POM)}
A computational model was developed to simulate the birefringence colors of the line, helix, ring, crossed helix, and twisted helix cell walls under POM. The program runs on GNU Octave.

\subsubsection{Transverse cuts}

\begin{figure*}
\centering
\includegraphics[width=1.0\textwidth]{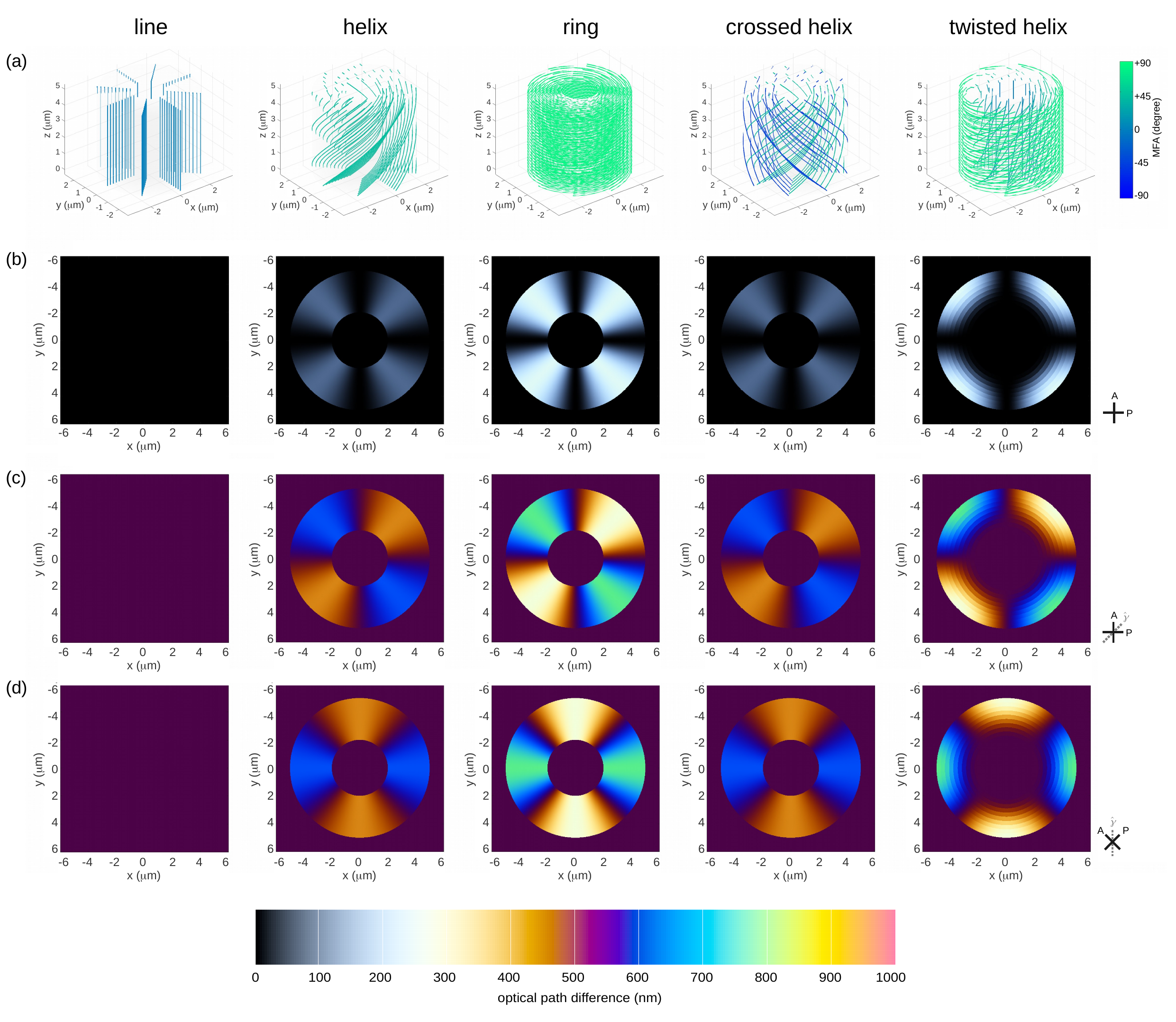}
\caption{Simulated POM images of the transverse cross sections of xylem cells.
A, analyzer.
P, polarizer.
$\gamma$, slow axis of the 530~nm retardation plate.
(a) 3D model of fibril structures. MFA is color-coded.
POM image of sample placed
between crossed P and A (b),
with P-A-$\gamma$= $0^\circ$-$90^\circ$-$45^\circ$ (c),
and with P-A-$\gamma$= $45^\circ$-$135^\circ$-$90^\circ$ (d).
Parameters for simulation: 
$n_o$, 1.528.
$n_e$, 1.578.
Thickness, 5~\textmu m.
Inner radius, 2~\textmu m.
Outer radius, 5~\textmu m.
10 layers of cylindrical shell.
A Birefringence color chart is attached below the POM images.}
\label{fig:Xylem_MFA_5modes_opt_top}
\end{figure*}

A xylem transverse cross section was modeled as a hollow cylinder (Fig.~\ref{fig:Xylem_MFA_5modes_opt_top}(a)), consisting of 10 layers of cylindrical shells. Each shell is a lamina of parallel cellulose fibrils with a specific MFA. The fibrils have right-handed twist.
The birefringence of wood pulp is in the range of 0.05$\pm$0.01~\cite{Delta_n_wood}.
The average refractive index of our wood was 1.528, measured with the Becke line test.
For the simulation, the ordinary ($n_o$) and extraordinary ($n_e$) refractive indices were set to 1.528 and 1.578, respectively.
The height of the cylinder was 5~\textmu m. The inner and outer radius of the cylinder were 2~\textmu m and 5~\textmu m, respectively.

Under the microscope, the light propagates along the axis of the cylinder ($\hat z$). The angle between $\hat k$ and the fibril axis, which is exactly the MFA and $\alpha$ in Eq.~\ref{eq:n_eff}, determines the birefringence $\Delta n$. The OPD is $\Delta n$ multiplied by the height of the cylinder, which is the thickness of the xylem (d).
OPD increases with increasing MFA, and thus line, helix, and ring fibril structures give zero, moderate, and maximum OPD, respectively.
Fibrils of $+45^\circ$ and $-45^\circ$ MFA in crossed helical structure show the same OPD.
Twisted helical structure has MFA from $0^\circ$ to $90^\circ$ from the innermost to the outermost layers, showing OPD from 0 to the maximum magnitude.

In crossed linear polarizer (P) and analyzer (A), the xylem slices show birefringence colors (Fig.~\ref{fig:Xylem_MFA_5modes_opt_top}(b)).
With accurate OPD, $\Delta n$ and thus $\alpha$ can be derived.
The azimuthal orientation of the fibril affects the transmittance.
When $\beta$ is $0^\circ$ or $90^\circ$, the linearly polarized input light remains the same polarization state and is subsequently blocked by the analyzer (A), showing a dark optical texture.
When the fibers are vertically upright on $\hat z$ ($\alpha$ is 0 and $\Delta n$ is 0), the zero OPD gives a dark texture, too.
Therefore the line structure is all dark, and the vortex-shaped helical structures show the four dark brushes of a Maltese cross.

To distinguish the fibrils pointing along $45^{\circ}$ or $135^{\circ}$, a retardation wave plate with 530~nm of retardation and the slow axis ($\hat \gamma$) on $45^{\circ}$ is placed between the sample and the A.
The fibrils on $45^{\circ}$ ($135^{\circ}$) have the 530~nm retardation added (subtracted) to the original OPD, so the $45^{\circ}$ or $135^{\circ}$ shows different colors.
As shown in Fig.~\ref{fig:Xylem_MFA_5modes_opt_top}(c), the magenta is due to the 530~nm retardation. The yellow at the 1st and 3rd quadrants and the blue at the 2nd and 4th quadrants represent a vortex-like fibril distribution. In case of lines, there is only ordinary mode of passing light, so the optical texture is all magenta.

The vertical fibril, the fibril on $0^\circ$, and the fibril on $90^\circ$ happen to show the same magenta color.
Rotating the P-A-$\gamma$ helps to distinguish the vertical from the in-plane cases. The vertical fibrils always contribute no OPD, and the birefringence color remains magenta. The in-plane fibrils show various brightness when P-A-$\gamma$ is rotating. In the following figures, we always show two typical cases:
the P-A-$\gamma =$ $0^\circ$-$90^\circ$-$45^\circ$ case (Fig.~\ref{fig:Xylem_MFA_5modes_opt_top}(c))
and 
the P-A-$\gamma =$ $45^\circ$-$135^\circ$-$90^\circ$ case (Fig.~\ref{fig:Xylem_MFA_5modes_opt_top}(d)).

\subsubsection{Longitudinal cuts}

\begin{figure*}
\centering
\includegraphics[width=1.0\textwidth]{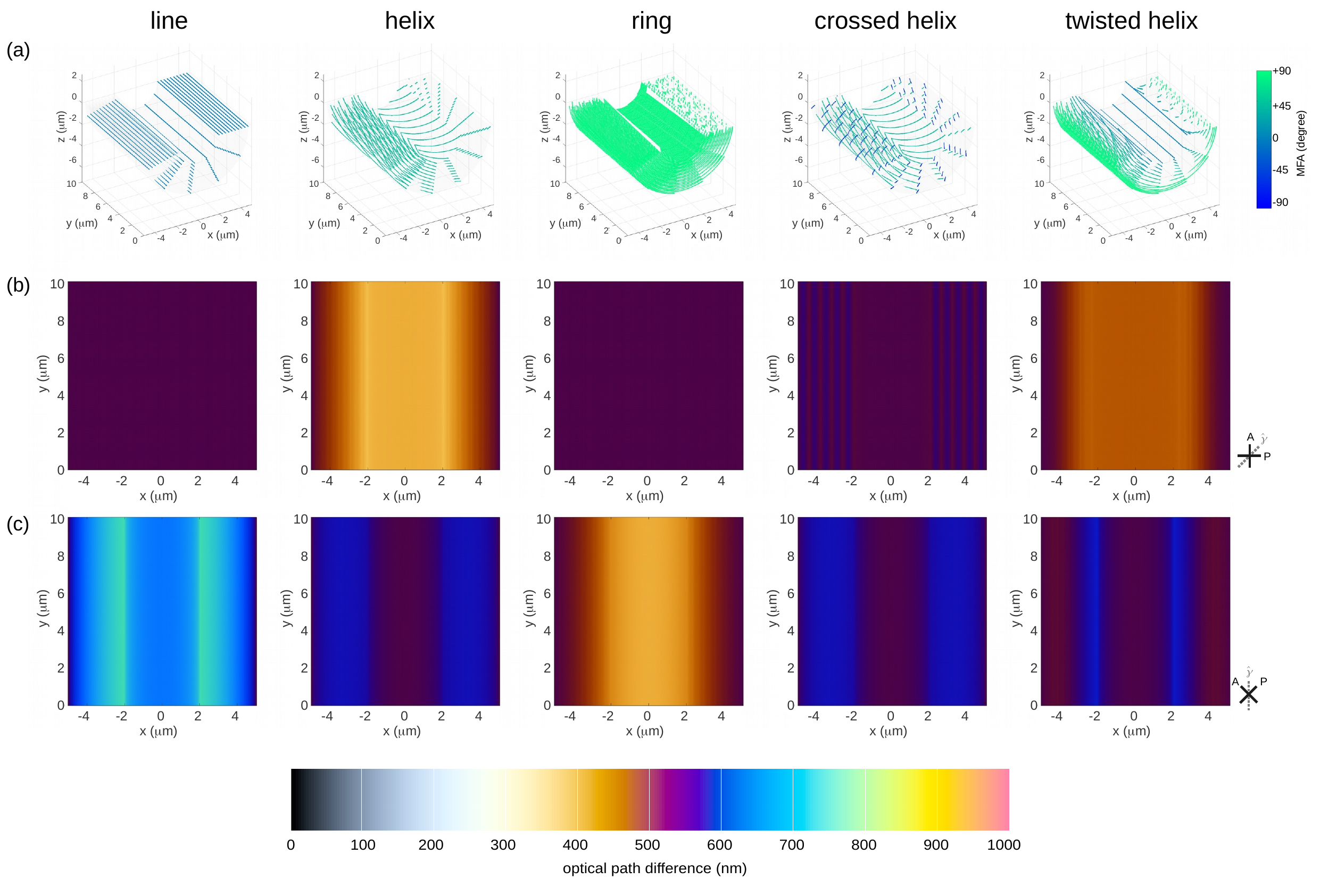}
\caption{Simulated POM images of the longitudinal cross sections of xylem cells.
A, analyzer.
P, polarizer.
$\gamma$, slow axis of the 530~nm retardation plate.
(a) 3D model of fibril structures. MFA is color-coded.
POM images
with P-A-$\gamma$= $0^\circ$-$90^\circ$-$45^\circ$ (b),
with P-A-$\gamma$= $45^\circ$-$135^\circ$-$90^\circ$ (c).
Parameters for simulation: 
$n_o$, 1.528.
$n_e$, 1.578.
Inner radius, 2~\textmu m.
Outer radius, 5~\textmu m.
10 layers of cylindrical shell.
A Birefringence color chart is attached below the POM images.
}
\label{fig:Fig_Xylem_MFA_5modes_opt_cut_half}
\end{figure*}

The longitudinal cross section of the cell wall is one-half of a hollow multi-layered cylinder (Fig.~\ref{fig:Fig_Xylem_MFA_5modes_opt_cut_half}(a)).
The objective lens of the POM sees the rectangular hollow section (RHS) of the cylinder.
In the simulation, we set the cross section in the x-y plane, and the half cylinder is in the range where z is negative. Light travels along the z-axis, passing through the half cylinder from -z to +z.
The computational model first calculates the $\Delta n$ and thickness of each cellulose lamina layer.
Along the light path, the direction of the optical axis changes layer by layer. Therefore, the total OPD of the entire cell wall must be the vectorial summation of OPD contributed by each layer of cellulose.
The characteristic traits are listed below:

\begin{itemize}

\item $\hat c$ in parallel to $\hat r$ is the additive position, where the magenta shifts to the blue.
\item $\hat c$ perpendicular to $\hat r$ is the subtractive position, where the magenta shifts to the yellow.

\item \textbf{Line}

P-A-$\gamma =$ $0^\circ$-$90^\circ$-$45^\circ$: All magenta.

P-A-$\gamma =$ $45^\circ$-$135^\circ$-$90^\circ$: Additive-position blue. The maximum OPD on the inner edge of the rectangular cross sections.

\item \textbf{Helix}

P-A-$\gamma =$ $0^\circ$-$90^\circ$-$45^\circ$: Subtractive-position yellow at the center curvy part.

P-A-$\gamma =$ $45^\circ$-$135^\circ$-$90^\circ$: Magenta at the center curvy part. Additive-position blue on the rectangular cross sections.

\item \textbf{Ring}

P-A-$\gamma =$ $0^\circ$-$90^\circ$-$45^\circ$: All magenta.

P-A-$\gamma =$ $45^\circ$-$135^\circ$-$90^\circ$: Subtractive-position yellow at the center curvy part.

\item \textbf{Crossed helix}

P-A-$\gamma =$ $0^\circ$-$90^\circ$-$45^\circ$: Magenta at the center. The rectangles show alternative yellow and blue strips.

P-A-$\gamma =$ $45^\circ$-$135^\circ$-$90^\circ$: Magenta at the center. The rectangles turn additive-position blue.

\item \textbf{Twisted helix}

P-A-$\gamma =$ $0^\circ$-$90^\circ$-$45^\circ$: Subtractive-position yellow at the center curvy part.

P-A-$\gamma =$ $45^\circ$-$135^\circ$-$90^\circ$: Mostly magenta except the additive blue on the inner side of the rectangles.

\end{itemize}

The longitudinal cross section optical texture is crucial for identifying the cellulose fibril structure.
Each fibril structure has its own distinct, unique color distribution (Fig.~\ref{fig:Fig_Xylem_MFA_5modes_opt_cut_half}(b) and (c)). By rotating the P-A-$\gamma$, the type of the cellulose organization can be easily and clearly recognized.

\section{\label{sec:results}Experimental results}

The trunk of an one-year-old \textit{Eucalyptus grandis} is sliced with sharp blades in the transverse and longitudinal directions.
The slices soaked in water or glycerol provide the best transmittance and resolution, while the matching refractive indices eliminate the fringe diffraction.
Thickness between 10~\textmu m and 20~\textmu m show the clearest birefringence color distribution.

\subsection{Large scale POM pictures}

\begin{figure*}
\centering
\includegraphics[width=1.0\textwidth]{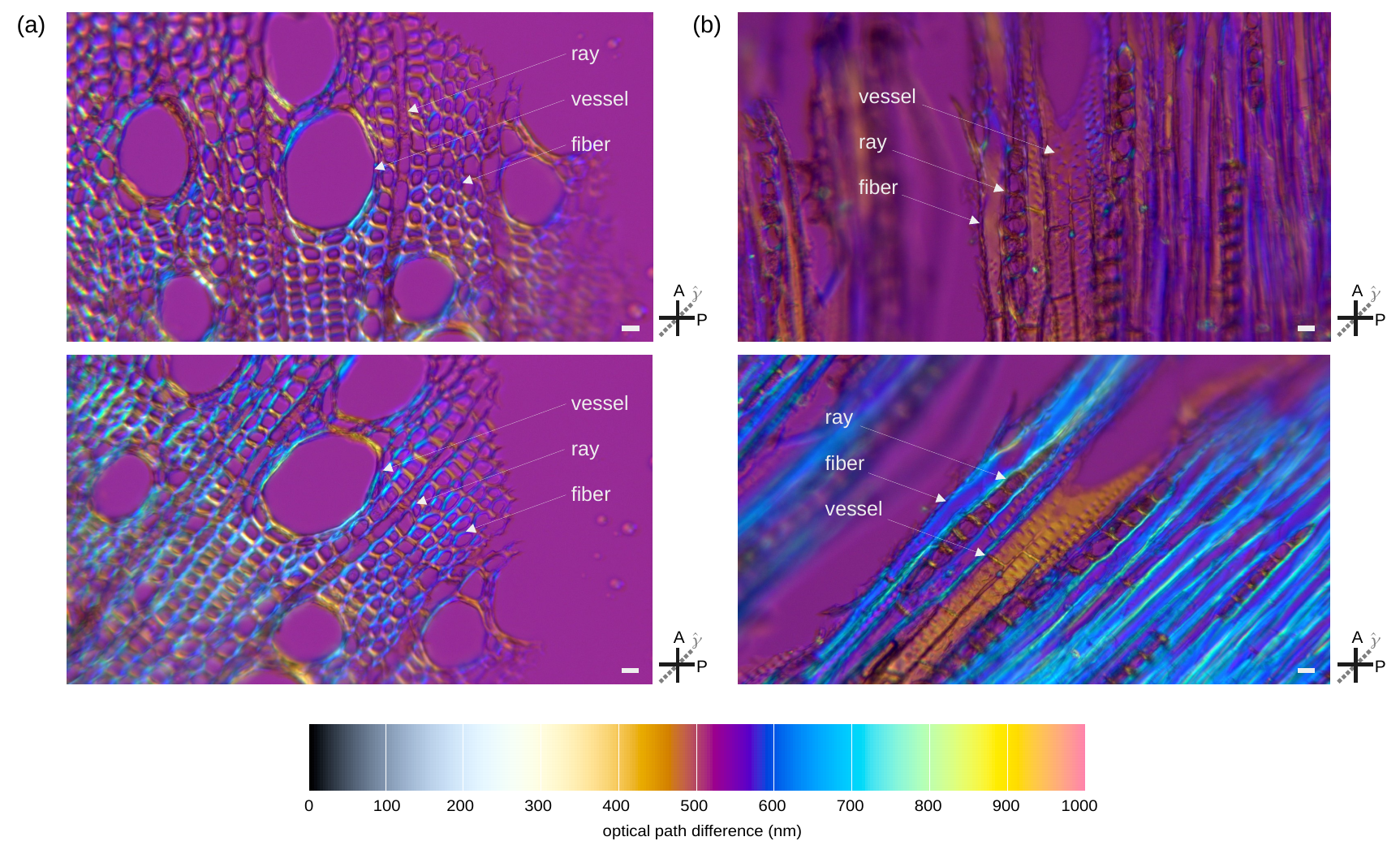}
\caption{Slices of xylem tissue under POM.
(a) Transverse cross section.
(b) Longitudinal cross section.
A, analyzer.
P, polarizer.
$\gamma$, slow axis of the 530 nm retardation plate.
Thickness of the slices ranges from 5 to 15~\textmu m.
Scale bare, 10~\textmu m.
A birefringence color chart is attached below the pictures.
}
\label{fig:POM_wood}
\end{figure*}

The transverse and longitudinal slices of xylem under POM are shown in Fig.~\ref{fig:POM_wood}(a) and Fig.~\ref{fig:POM_wood}(b), respectively. The large rings in Fig.~\ref{fig:POM_wood}(a) are the circular hollow sections (CHS) of the vessel cell. The small, densely packed rings are the fiber cells. Between the rings, the long straight thin walls are the RHS of the ray cells.
The CHS of the vessel cell shows the largest OPD among all the cells, which is $\approx$ 170~nm. The yellow (cyan) in the 1st and 3rd (2nd and 4th) quadrants is the birefringence color of  530+170 nm (530-170 nm) OPD.
The fiber cells show OPD of $\approx$ 50~nm (530+50 nm, the navy. 530-50 nm, the orange.)
Overall, the $\Delta n$ and thus MFA of a vessel cell is the largest among the three.

In Fig.~\ref{fig:POM_wood}(b), the huge tube with pits on the wall is the vessel, and the long, straight, densely packed tubes are the fiber cells. The rings between the straight cell walls are the transverse cross section of ray cells.
When the cylinder axes are on $90^\circ$ and $\gamma$ on $45^\circ$ (Fig.~\ref{fig:POM_wood}(b), top panel), most of the cells show magenta of little OPD, except the RHS of the fiber cells is slightly yellow in the subtractive position.
When the tube axes are on $45^\circ$, aligned with $\gamma$ (Fig.~\ref{fig:POM_wood}(b), bottom panel), the image shows significant birefringence colors. The RHS of the vessel shows yellow of 530-100 nm OPD, meaning that the fiber is on $135^\circ$.
The center curvy parts of the fiber cells are dark navy of 530+30 nm OPD. The thin walls are cyan of 530+170 nm OPD, meaning that the fibrils are in parallel with $\gamma$.
The birefringence color clearly shows that the vessels and fiber cells have different fibril structures.

\subsection{One cell: transverse cross section}

\begin{figure*}[ht!]
\centering
\includegraphics[width=0.7\textwidth]{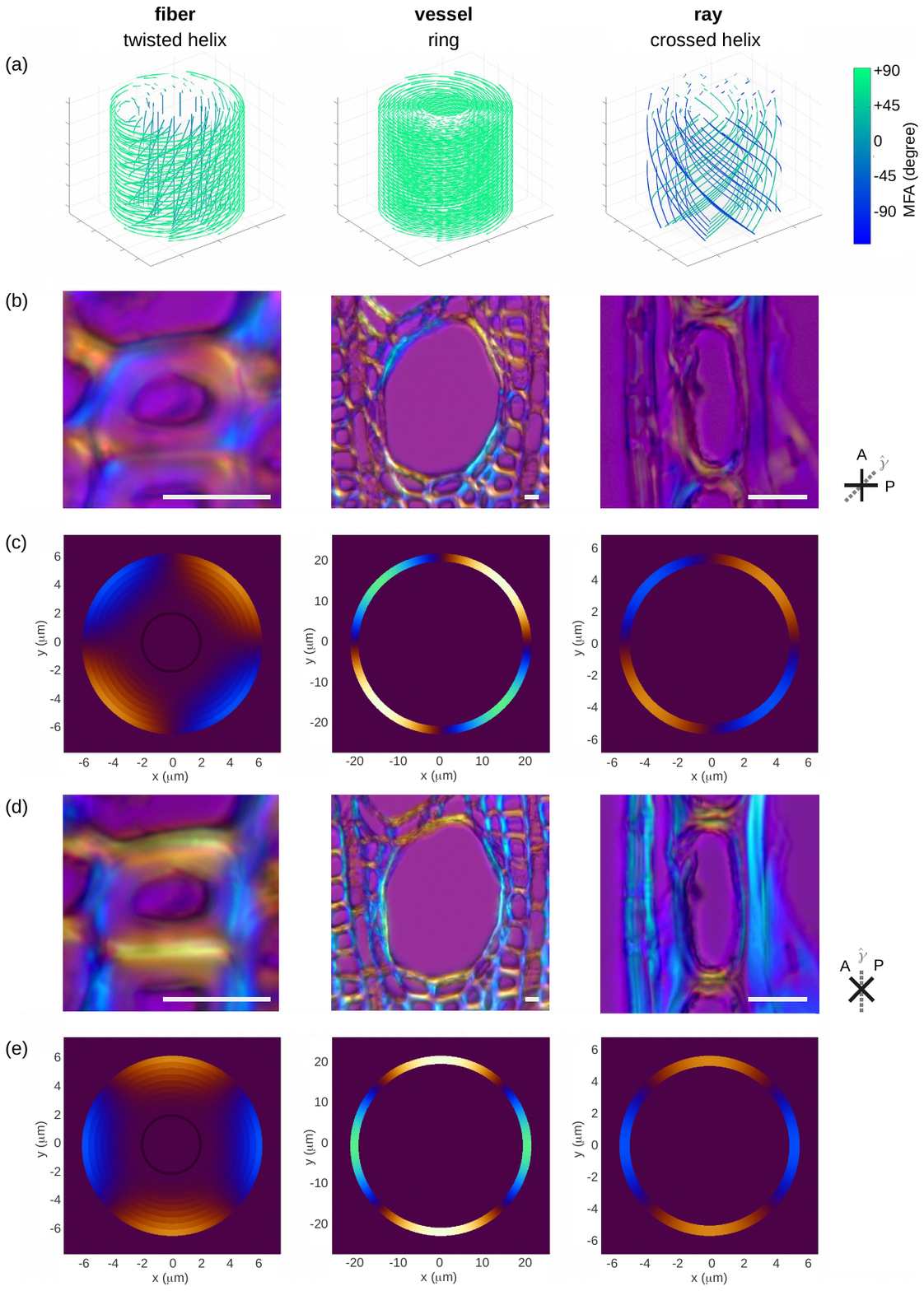}
\caption{Transverse cross sections of fiber, vessel, and ray cells under POM.
A, analyzer.
P, polarizer.
$\gamma$, slow axis of the 530~nm retardation plate.
(a) 3D fibril structures. MFA is color-coded.
(b) POM images of the single cells with P-A-$\gamma$= $0^\circ$-$90^\circ$-$45^\circ$.
(c) Simulated POM images of the same cells in (b). 
(d) POM images of the single cells with P-A-$\gamma$= $45^\circ$-$135^\circ$-$90^\circ$.
(e) Simulated POM images of the same cells in (d). 
Thickness of fiber cells, 10~\textmu m. Vessel cell, 8~\textmu m. Ray cell, 8~\textmu m.
Scale bar, 10~\textmu m.
Parameters for the simulation: 
$n_o$, 1.528.
$n_e$, 1.578.
Thickness, 5~\textmu m.
MFA of the twisted helix is from 0$^\circ$ to 45$^\circ$. The vessel, 90$^\circ$. The crossed helix, $\pm$45$^\circ$.
}
\label{fig:POM_wood_zoom_ring}
\end{figure*}

The POM images of the transverse slices of the fiber, vessel, and ray cells are shown in Fig.~\ref{fig:POM_wood_zoom_ring}.
The simulated birefringence color distributions of the circular hollow sections (CHS) are placed below the POM images.
Fig.~\ref{fig:POM_wood_zoom_ring}(a) shows the 3D schematic diagram of the cellulose fibrils.
Fig.~\ref{fig:POM_wood_zoom_ring}(b) and (c) show the experimental and simulated POM images with
P-A-$\gamma =$ $0^\circ$-$90^\circ$-$45^\circ$, respectively.
Fig.~\ref{fig:POM_wood_zoom_ring}(d) and (e) show the pictures with P-A-$\gamma =$ $45^\circ$-$135^\circ$-$90^\circ$.
The fiber cells show magenta around the hole in the cell (lumen), and the magenta shifts toward orange (the first and third quadrats) and navy (the 2nd and 4th quadrat) with four magenta brushes from the inner to the outer side of the wall.
The magenta around the hole remains the same while the P-A-$\gamma$ rotates, but the orange-navy texture rotates with the P-A-$\gamma$. This verifies that the cellulose fibril is vertical (MFA=$0^\circ$) around the lumen of the cell, and the fibrils twist away, turning into a helical organization.
The vessel and ray cells shows rather evenly distributed colors.
There is no magenta color around the lumen, so they seem to be helical.
Vessel shows the largest OPD and thus the largest MFA among the three.

\subsection{One cell: longitudinal cross section}

\begin{figure*}[ht!]
\centering
\includegraphics[width=0.7\textwidth]{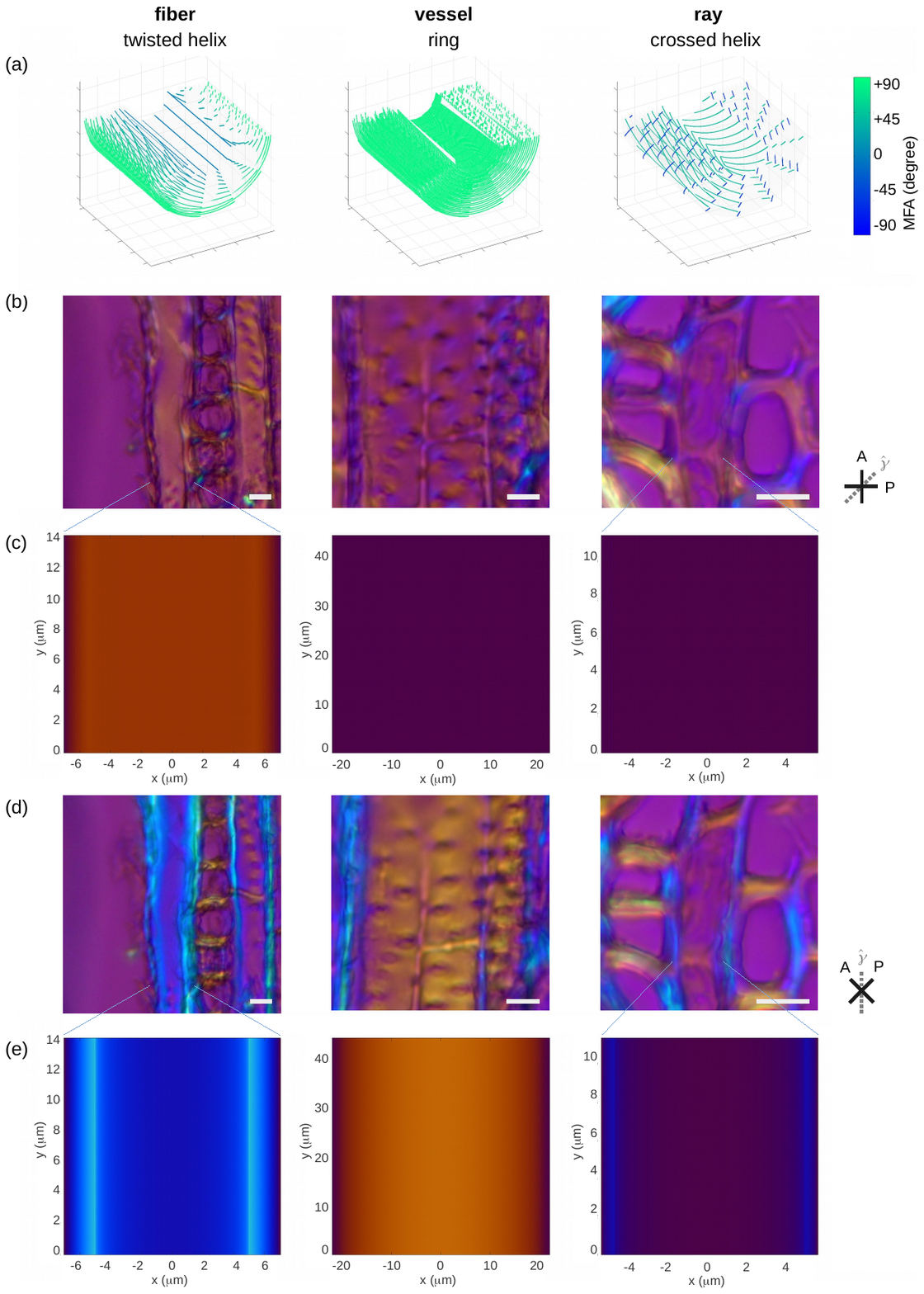}
\caption{Longitudinal cross sections of fiber, vessel, and ray cells under POM.
A, analyzer.
P, polarizer.
$\gamma$, slow axis of the 530~nm retardation plate.
(a) 3D fibril structures. MFA is color-coded.
(b) POM images of the single cells with P-A-$\gamma$= $0^\circ$-$90^\circ$-$45^\circ$.
(c) Simulated POM images of the same cells in (b). 
(d) POM images of the single cells with P-A-$\gamma$= $45^\circ$-$135^\circ$-$90^\circ$.
(e) Simulated POM images of the same cells in (d). 
Thickness of fiber cells, 8~\textmu m. Vessel cell, 25~\textmu m. Ray cell, 6~\textmu m.
Scale bar, 10~\textmu m.
Parameters for the simulation: 
$n_o$, 1.528.
$n_e$, 1.578.
Cell wall thickness of the fiber cell, 4.25~\textmu m. Vessel cell, 2~\textmu m. Ray cell, 0.65~\textmu m.
MFA of the twisted helix is from 0$^\circ$ to 45$^\circ$. The vessel, 90$^\circ$. The crossed helix, $\pm$45$^\circ$.}
\label{fig:POM_wood_zoom_rectangle}
\end{figure*}

The longitudinal slices of the the fiber, vessel, and ray cells are shown in Fig.~\ref{fig:POM_wood_zoom_rectangle}.
The RHSs were observed with the P-A-$\gamma =$ $0^\circ$-$90^\circ$-$45^\circ$ and P-A-$\gamma =$ $45^\circ$-$135^\circ$-$90^\circ$ settings. Simulated birefringence color distribution is right below each POM picture.
When the P-A-$\gamma$ is rotating, the semi-cylinder of the fiber cell changes color from orange to dark navy, and the cell wall cross sections turn from magenta to cyan. This is exactly the optical texture of a twisted helical semi-cylinder, as predicted by the simulation.
The vessel cell wall with pits is mostly magenta with P-A-$\gamma =$ $0^\circ$-$90^\circ$-$45^\circ$. Around the pits, the orange and navy reveal the fibrillar vortices. With P-A-$\gamma =$ $0^\circ$-$90^\circ$-$45^\circ$, the yellow of subtractive position clearly show that the cellulose fibrils are normal to $\gamma$, meaning that this vessel has the ring structure.
The semi-cylinder part of ray cell remains magenta even if the  P-A-$\gamma$ is rotating.
Only the cell wall RHS parts turn slightly blue in the additive position when $\hat \gamma$ is $90^\circ$. Most likely, the cellulose fibrils in ray cells are crossed, so the optical retardation contributed by each layer of cellulose eventually cancels out. 

Based on the results presented in Fig.~\ref{fig:POM_wood_zoom_ring} and Fig.~\ref{fig:POM_wood_zoom_rectangle}, the cellulose fibril structures of fiber, vessel, and ray cells are twisted helix, rings, and crossed helix, respectively.

\subsection{Measuring microfibril angles (MFA)}

\begin{figure*}
\centering
\includegraphics[width=1.0\textwidth]{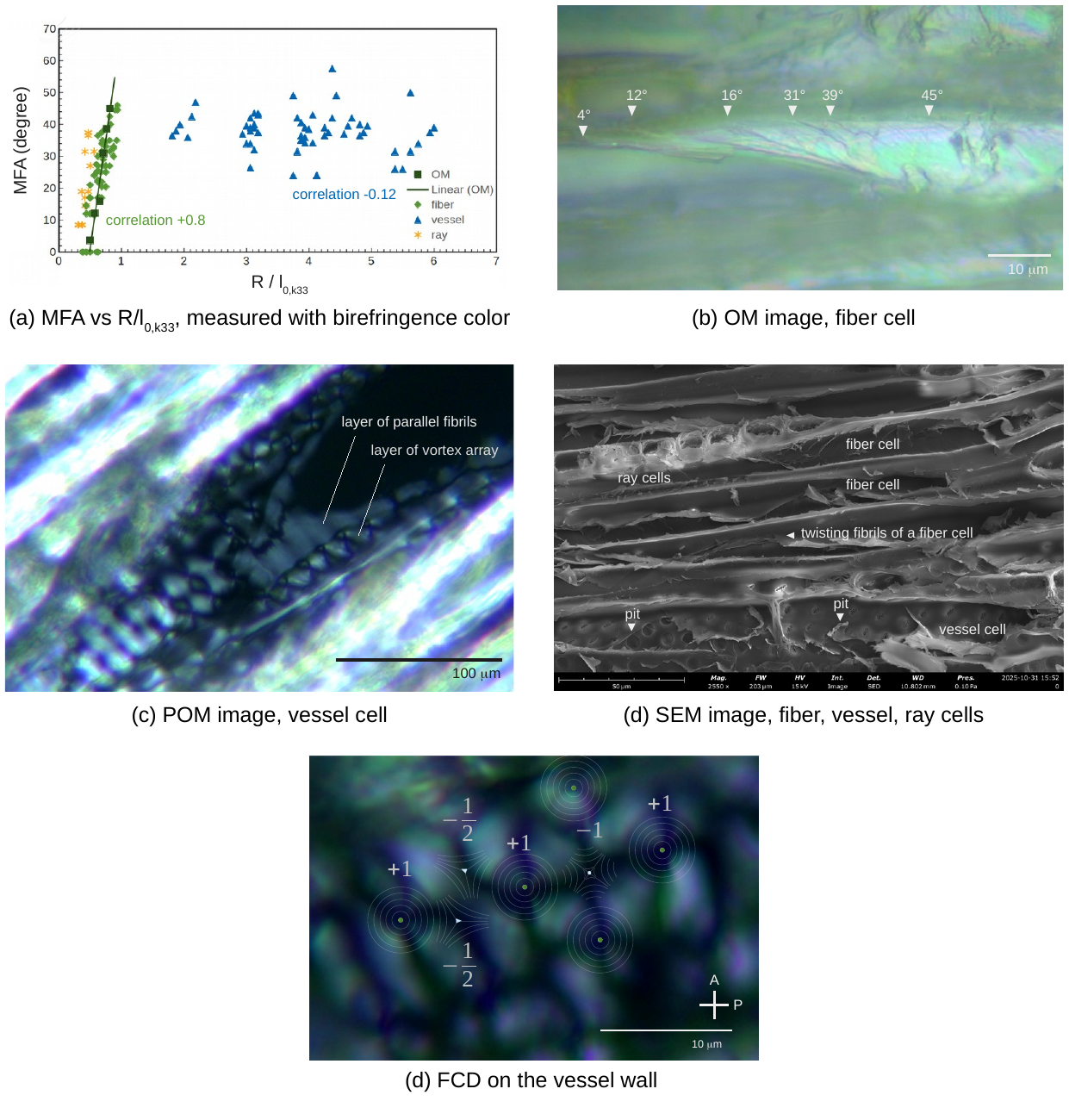}
\caption{\label{fig:Xylem_MFA}Measuring the MFA. 
(a) MFA versus $\frac{R}{l_0, k_{33}}$ measured with POM birefringence colors.
Square markers show the MFA presented in the OM images.
Diamond markers are the MFA measured with birefringence color spectrum.
Vessel cell data, triangles.
Ray cell data, stars.
(b) Reflection OM image of a fiber cell.
(c) POM image of a longitudinal cross section of a vessel cell in crossed polarizer and analyzer. The inset image show the director field of the fibrils and topological charge of the defects and disclinations. 
(d) SEM image of a longitudinal cross sections of fiber, vessel, and ray cells.
(e) Focal conic domains (FCD) on the vessel wall.}
\end{figure*}

The birefringence colors of the circular hallow cross-sections of the cell wall is directly related to the local MFA. The OPD is $\Delta n$ multiplied by the thickness of the slice d, and the $\Delta n$ is a function of MFA (Eq.~\ref{eq:n_eff}). The OPD is measured with the birefringence color spectrum. d is measured with the Fabry-P{\'e}rot interferometry. Then the MFA can be extracted from the POM images.
MFA versus $\frac{R}{l_{0, k_{33}}}$ of fiber, vessel, and ray cells are summarized in Fig.~\ref{fig:Xylem_MFA}(a).
Here we assume $l_0$ is $\frac{k_{33}}{\gamma}$ of carboxylated cellulose nanocrystal in water, which is 8~\textmu m, since it shows isotropic, nematic, and chiral liquid crystal phases, similar to the fibril organization in the cell wall.
MFA of fiber cell increases with respect to $\frac{R}{l_{0, k_{33}}}$ from $0^\circ$ to $45^\circ$ as $\frac{R}{l_{0, k_{33}}}$ increases from 0.4 to 0.9. The correlation coefficient is +0.8, suggesting a very strong association. The data from the OM picture (the square markers, picture in Fig.~\ref{fig:Xylem_MFA}(b)) and the ones measured with the POM birefringence color (the diamond markers) are in excellent agreement. 
The $\frac{R}{l_{0, k_{33}}}$ of vessel cell (marked by the triangles) span a large range from 1.8 to 6.0. The MFA scatters between $22^\circ$ and $59^\circ$, barely correlated with $\frac{R}{l_{0, k_{33}}}$ (correlation coefficient -0.12).

Optical and electron microscope pictures of crushed cell walls show the fibril bundles, providing the reliable references.
The wood tissue was crushed in a way that the laminae split along the fibril axis, so the MFA is clear in the picture. 
A tubular fiber cell wall has six layers of laminae (Fig.~\ref{fig:Xylem_MFA}(b)), The MFA are $4^\circ$, $12^\circ$, $16^\circ$, $31^\circ$, $39^\circ$, $45^\circ$, from the inner to the outer layers, respectively.
The twist is right-handed.
A vessel cell wall is composed of a well-aligned fibril layer overlaid with a layer of porous tissue (Fig.~\ref{fig:Xylem_MFA}(c)). We noticed that the MFA of the well aligned fibrils varies from cell to cell. They can be rings or helix. In the porous layer, the fibrils circle around the pores (the pits). Therefore the optical texture in Fig.~\ref{fig:POM_wood}(b) is actually rendered by one layer of azimuthal fibrils and one layer of vortex arrays.
The well aligned layer provides the whole birefringent color. The little vortices contribute the local Maltese cross. The total OPD should be the vectorial summation of the two compositions. The scattered MFA of vessel cell in Fig.~\ref{fig:Xylem_MFA}(a) is expected because the fibrils circling around the pits can contribute a wide range of angles.
The SEM picture (Fig. ~\ref{fig:Xylem_MFA}(d)) shows the ray, fiber and vessel cells. The ray cells are composed of layers of thin laminae, which is so thin that the fibril direction cannot be recognized. The longitudinal cuts of the fiber cells show the fibrils splitting. A particular cell reveals its twisting plywood structure. The vessel cell shows the fibrils circling around the pits. Note that the vessel cell wall can split vertically, horizontally, or in diagonal direction, agreeing with the vortex texture observed in Fig.~\ref{fig:Xylem_MFA}(c).

The ray cell wall is so thin that the microscope cannot resolve the layers of the fibrils. Under the POM, ray cell wall is generally composed of two sections of thin film. The one nearby the lumen has axial fibrils (MFA between $9^\circ$ and $18^\circ$), and the outer layer is likely a helix (MFA between $28^\circ$ and $38^\circ$).

\section{Discussion}

\subsection{Fiber cell: twisted helix}

The fiber cell possesses the twisted helical fibrils, the Bouligand twisted plywood structure, which has been found in stone cells and seed coats. The twisted helix is stiff and hard to break, offering good support and protection for the soft vessel cells and delicate ray cells.

Twisted plywood of fibrils can be found in animal bones, beetles' shells, and the claws of arthropods~\cite{CLC_living_Mitov}. The bones are bundles of osteons, which are tubes of twisted helical collagen fibers. The fiber cells are just like the bone of a tree. The twisted plywood can be made of cellulose, collagen, or chitin, suggesting that the formation of twisted plywood is a physical process. The genes and chemical reactions determine the chirality of the molecules. Then the lyotropic liquid crystal phase transition (isotropic $\Leftrightarrow$ nematic $\Leftrightarrow$ chiral) and the pattern formation in curved confinements lead to the formation of bones and shells.

Curvature-induced fibril alignment is critical to the twist of fibrils, since the MFA is highly correlated with the curvature of the cell wall (Fig.~\ref{fig:Xylem_MFA}(a)). Rigid, inflexible fibrils tend to be vertical in narrow tubes to reduce deformation ($\frac{R}{l_{0,k_{33}}} < 0.4$). As the radius increases, the fibrils can tilt to form a helical structure for the dense packing ($0.4 < \frac{R}{l_{0,k_{33}}} < 0.9$). Once $\frac{R}{l_{0,k_{33}}}$ exceeds 2, the MFA becomes independent of the tube radius, meaning that the cell wall seems flat for the cellulose fibrils and the directional anchoring is weak.
The curvature causes bending ($k_{33}$). The restoring force from the bending competes with the anchoring on the surface ($\gamma \omega_{\phi}$). $\frac{k_{33}}{\gamma \omega_{\phi}}$ can be the internal length scale $l_0$, measuring the effectiveness of the curvature induced alignment.

The MFA versus $\frac{R}{l_0, k_{33}}$ increases linearly, meaning that the fibrils can be at the chiral liquid crystal state. The MFA increases from $0^\circ$ to $45^\circ$ in 4~\textmu m, so the pitch is about 16~\textmu m. The pitch seems much larger than the thickness of a fiber cell wall, and therefore the confinement dominates the fibril alignment. 

\subsection{Vessel cell: helix + vortex array}

Vessel cell wall is composed of a layer of helical fibrils and a layer of pores. The MFA of the helical layer varies from cell to cell between $45^\circ$ to $90^\circ$. The pores are vortices of fibrils, arranged in arrays.

When a curved confinement, antagonistic boundary or phase transition forces chiral liquid crystal (CLC) to bend, it bends into concentric shapes in order to balance the tension from the antagonistic anchoring and the intrinsic twist.
The concentric structures pack into hexagonal or square arrays, called the focal conic domains (FCDs).
This has been found on the beetles' shells~\cite{FCD_beetles}, CLC droplets~\cite{FCD_droplets_Randal}, and CLC shells~\cite{FCD_shells_Tran, FCD_shells_Lavrentovich, FCD_shells_Lagerwall}.

Under POM, the concentric vortex part shows the four-brush Maltese cross with a topological charge of +1. Occasionally they split into a pair of +1/2 disclinations that attach on the two poles of the elliptical pore.
The interstitial region bounded by four neighboring vortices is hyperbolic hedgehog-shaped with a topological charge of -1. In a relaxed, loose array, the hyperbolic hedgehog split into two $-\frac{1}{2}$ disclinations.
In a hexagonal array, the area among the three neighboring vortices shows the $-\frac{1}{2}$ disclination. The area between the vortices can also be defect-free homogeneous alignment, which also presents in our vessel POM pictures.

The vortices and the mixture of hexagonal and square arrays is the key feature of CLC. Especially the spontaneous array is unique to focal conic domains on a curved homeotropic boundary. 
On the other hand, the wavy dark brushes, the splitting topological defects to minimize free energy, and particularly the half-integer disclinations are the characteristic of nematic order~\cite{NLC_pore_array, NLC_vortex}.
The vessel porous layer clearly presents as a liquid crystal undergoing phase transition between nematic and chiral order.

The formation of vessel cells proceeds at least in two stages: the first stage generates the curvature induced nematic helical order, and the second stage is the curvature induced focal conic domains of chiral liquid crystal. Between the two stages, the surface anchoring switches from in-plane to homeotropic vertical, and the cellulose fibrils turn chiral. The timing and the root cause of the dramatic shift will help answer how the xylem cells differentiate.

\subsection{Ray cell: crossed fibrils}

In case of ray cells, the cell wall is extremely thin and multi-layered. The only reliable information is that the fibrils are crossed. Current data is not able to determine the liquid crystal phase and the accurate MFA. 
The ray cell is the only living cell among the three. Possibly, the concentration of cellulose in ray cell is low. Therefore, the curvature-induced alignment, the surface anchoring, and the liquid crystal order are not significant in general in a living cell.

\subsection{POM uncovers the liquid crystal pattern}

In this research, the 3D fibril structure is clearly identified with the birefringent-colored optical texture under POM.
The distribution of MFA can be measured numerically, and the volume of data is sufficient for statistical analysis.
Additionally, the pattern of topological defects indicates the nematic and chiral order of cellulose.
Polarized optical microscopy offers abundant information holistically in one picture. The sample can stay in water or glycerol, and the sample preparation does not destroy the 3D structure.
Electron microscopy suffers a lot from the damaging sample preparation and the extremely localized view, making statistical analysis barely viable. Most important of all, electron microscopic pictures cannot distinguish the cellulose fibril from the lignin and hemicellulose matrix, causing lots of confusion and inconsistent results in the past decades.
Our results prove that POM is very sensitive to the birefringent cellulose, while the optically isotropic matrix does not interfere with the measurement.
Utilizing the interaction between polarized light and the cellulose effectively improves the 3D and statistical analysis of xylem cell walls.

\section{Conclusion}

Wood tissue comprises a network of twisted helical (fiber), vortex-array (vessel), and crossed helical (ray) cellulose tubes, demonstrated by our POM and complemented by OM and SEM pictures.
Crucially, the identification of isotropic, nematic, and chiral orders within the xylem cell wall strongly suggests that liquid crystal phase transitions drive the structural development of wood.
Our results show that the organization of cellulose in wood begins with geometric frustration and ends with an organized, self-retained, and functional 3D structure. Frustration has been the core challenge in modern condensed matter physics, as it reveals how a physical system resolve conflicts when its internal microscopic forces are fundamentally incompatible with the external boundary constrains. In case of wood, the anisotropic fibril liquid crystal adapts by forming helical tubes and vortex arrays, transforming the the brittle lamina into elastic, tear-resistant, and self-healing tissue capable of sustaining life. Out of the frustration, trees exploits topological laws to generate its own structural and fluidic networks.
The 3D microfibril model established herein can be applied to investigate how the helical channels and pore arrays improve capillary transport while minimizing the risk of cavitation, which benefits the development of efficient water purification, energy-saving water transport, and hydrovoltaic devices.
Understanding the microscopic 3D cellular network of wood yields design principles for superior mechanical reinforcement. It also helps to fabricate porous building materials to regulate moisture and thermal transfer, providing sustainable solutions for climate control.
Mimicking these native chiral architectures facilitates the fabrication of biomimetic scaffolds and meta materials for optical fibers and optoelectronics for generation, transmission, and storage of quantum information.
While this study focused exclusively on mature xylem cells, the precise onset and process of the cellulose liquid crystal phase transition remains to be elucidated. Future work will follow the chronological sequence of growth, spanning cambial initials, differentiating cells, cells undergoing secondary cell wall thickening, and fully mature cells. By systematically tracking the microstructural evolution across these developmental stages, we aim to map the trajectory of the liquid crystal phase transition. This will clarify the fundamental role of cellulose liquid crystals in driving and regulating plant growth.


\begin{acknowledgments}
We are grateful to Prof. Yu-Chieh Cheng, Dr. Wei-Ting Hsu and Prof. Chia-Chih Chang for the unconditional support on Scanning Electron Microscopy and chemistry. We are grateful to Prof. Ying-Chung Lin for gifting us the eucalyptus sapling. It had previously suffered from powdery mildew due to drought and irregular sunlight, resulting in a height of only one meter. This year, it decided to bypass its withered and twisted trunk, sprouting new branches from its base. Now it is a lush little tree, covered in vibrant green leaves, giving off its rosy refreshing scent.
This research was supported by the National Science and Technology Council, Taiwan, under Grant no. NSTC 111-2112-M-A49-042-MY3 and Grant no. NSTC 114-2112-M-A49-24. This work is supported by the Center for Emergent Functional Matter Science of National Yang Ming Chiao Tung University from The Featured Areas Research Center Program within the framework of the Higher Education Sprout Project by the Ministry of Education (MOE) in Taiwan.
\end{acknowledgments}

\nocite{*}

\bibliography{apssamp}
\end{document}